\documentclass[%
reprint,
superscriptaddress,
preprintnumbers,
nofootinbib,
amsmath,amssymb,amsthm,
aps,prl,
tikz,
]{revtex4-2}

\usepackage{isomath}
\usepackage{amsmath,amsthm}
\usepackage{amsbsy}
\usepackage{amssymb}
\usepackage{amscd}
\usepackage{amsfonts}
\usepackage{stmaryrd}
\usepackage{multirow}
\usepackage{booktabs}
\usepackage[utf8]{inputenc}
\usepackage[T1]{fontenc}
\usepackage{newtxtext} 
\usepackage{microtype}

\usepackage{graphicx}
\usepackage{caption}
\usepackage{subcaption}
\usepackage{boxedminipage}
\usepackage{calc}
\usepackage[dvipsnames]{xcolor}
\graphicspath{ {media/} }
\usepackage{tocloft}

\usepackage[normalem]{ulem}

\usepackage{orcidlink}

\usepackage{upgreek}

\theoremstyle{definition}

\AtEndEnvironment{definition}{\null\hfill\qedsymbol}

\AtEndEnvironment{remark}{\null\hfill\qedsymbol}

\AtEndEnvironment{example}{\null\hfill\qedsymbol}

\AtEndEnvironment{assumption}{\null\hfill\qedsymbol}

\usepackage{hyperref}

\begin{document}

%%%%%%%%%%%%%%%%%%%%%
%%%%%%%%%%%%%%%%%%%%%
%%%%%%%%%%%%%%%%%%%%%
%%%%%%%%%%%%%%%%%%%%%

\preprint{To appear in Physical Review Letters (DOI: \href{https://doi.org/10.1103/j8x1-hy4t}{10.1103/j8x1-hy4t} )}

\title{Compressive Splitting in Brittle Solids: The Inverse of Wrinkling in Sheets}

\author{Maryam Khodadad\orcidlink{0009-0000-9284-4910}}
    \email{mkhodada@andrew.cmu.edu}
    \affiliation{Program in Computational Mechanics, Carnegie Mellon University}
    \affiliation{Department of Civil and Environmental Engineering, Carnegie Mellon University}

\author{Francois Barthelat}
    \affiliation{Department of Mechanical Engineering, University of Colorado}

\author{John D. Clayton\orcidlink{0000-0003-4107-6282}}
    \affiliation{Terminal Effects Division, Army Research Laboratory}

\author{George Gazonas\orcidlink{0000-0002-2715-016X}}
    \affiliation{Biology, Materials and Manufacturing Sciences Division, Army Research Laboratory}

\author{Kaushik Dayal\orcidlink{0000-0002-0516-3066}}
    \email{Kaushik.Dayal@cmu.edu}
    \affiliation{Department of Civil and Environmental Engineering, Carnegie Mellon University}
    \affiliation{Center for Nonlinear Analysis, Department of Mathematical Sciences, Carnegie Mellon University}
    \affiliation{Department of Mechanical Engineering, Carnegie Mellon University}

\date{\today}

%%%%%%%%%%%%%%%%%%%%%
%%%%%%%%%%%%%%%%%%%%%
%%%%%%%%%%%%%%%%%%%%%
%%%%%%%%%%%%%%%%%%%%%

\begin{abstract}
    Axial splitting is the dominant failure mode of brittle solids under compression, yet its mechanical origin remains unclear. 
    We show that clamped loading platens suppress lateral Poisson expansion, generating boundary-induced tensile stresses at the specimen interior -- the compressive analog of wrinkling in stretched sheets. 
    This mechanism provides a predictive strength law, relating axial splitting to tensile strength, geometry, and confinement pressure.
    This is validated against diverse materials ranging from rocks to ceramics, establishing axial splitting as a geometry-controlled elastic process rather than a stochastic flaw problem.
\end{abstract}

\maketitle

%%%%%%%%%%%%%%%%%%%%%
%%%%%%%%%%%%%%%%%%%%%
%%%%%%%%%%%%%%%%%%%%%
%%%%%%%%%%%%%%%%%%%%%
The compressive failure of brittle solids is a fundamental problem in mechanics with broad implications for geophysics, civil engineering, and materials science. Under uniaxial or low-confinement compression, these materials fail predominantly via axial splitting—the formation and propagation of fractures parallel to the loading axis. This phenomenon is strikingly universal across microstructurally diverse systems, including rocks~\cite{li2002progressive, yang2016experimental}, concrete~\cite{bazant1998fracture}, high-performance ceramics~\cite{chen2015experiments, subhash1995dynamic}, and ice~\cite{schulson1997across, bohm2022data}. The observation of axial splitting across such fundamentally different materials suggests a universal mechanism rooted in continuum mechanics, rather than one dictated solely by the statistics of microscopic flaws. 
Analogous to macroscopic instabilities emerging in diverse systems, from metallic alloys to granular media~\cite{Valdes2017PRL}, identifying a continuum-level deterministic driver for splitting is critical for predicting geological scale fracture and designing damage-tolerant structures.

\begin{figure}[ht!]
    \centering
    \begin{subfigure}[t]{0.35\linewidth}
        \centering
        \includegraphics[height=4.9cm]{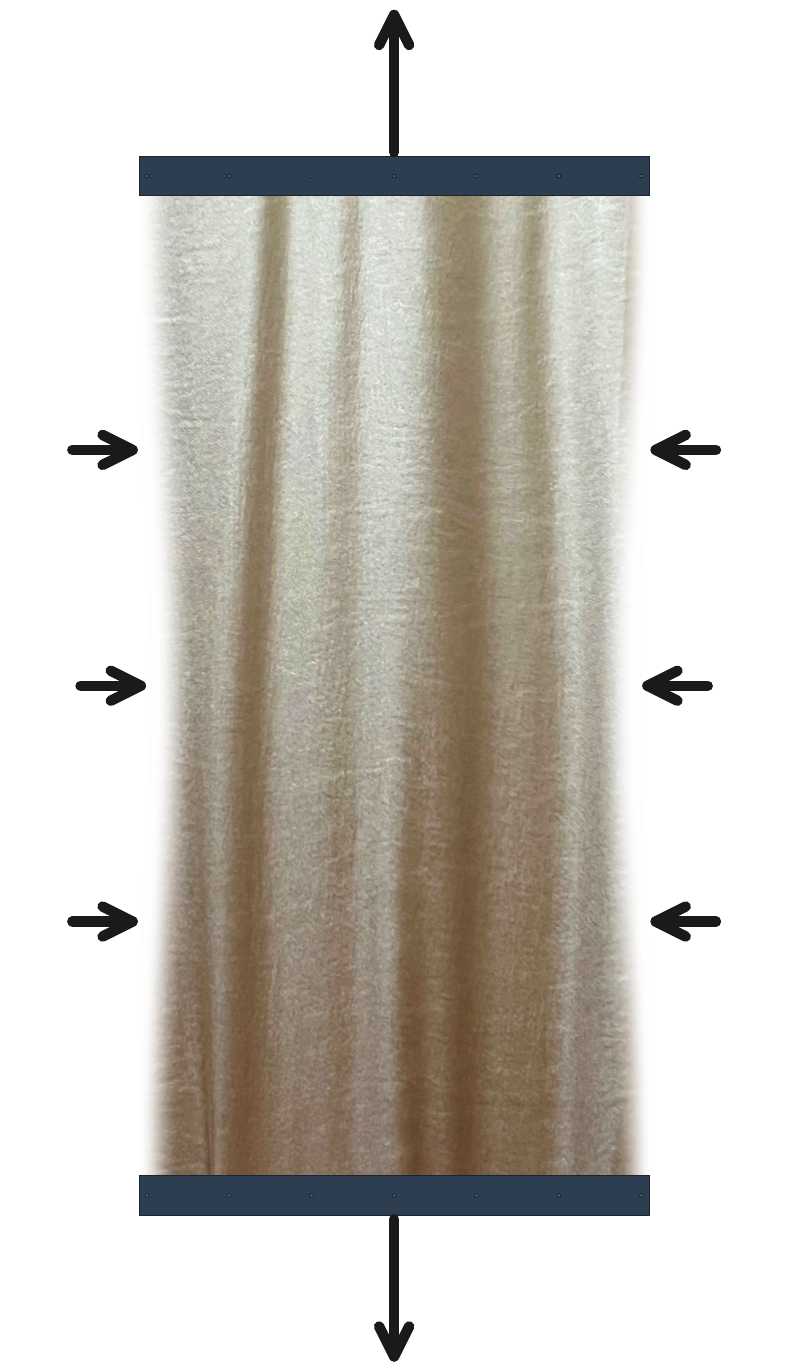}
        \caption{}
        \label{fig:schematic_wrinkling}
    \end{subfigure}
    \hspace{5mm}
    \begin{subfigure}[t]{0.35\linewidth}
        \centering
        \includegraphics[height=5cm]{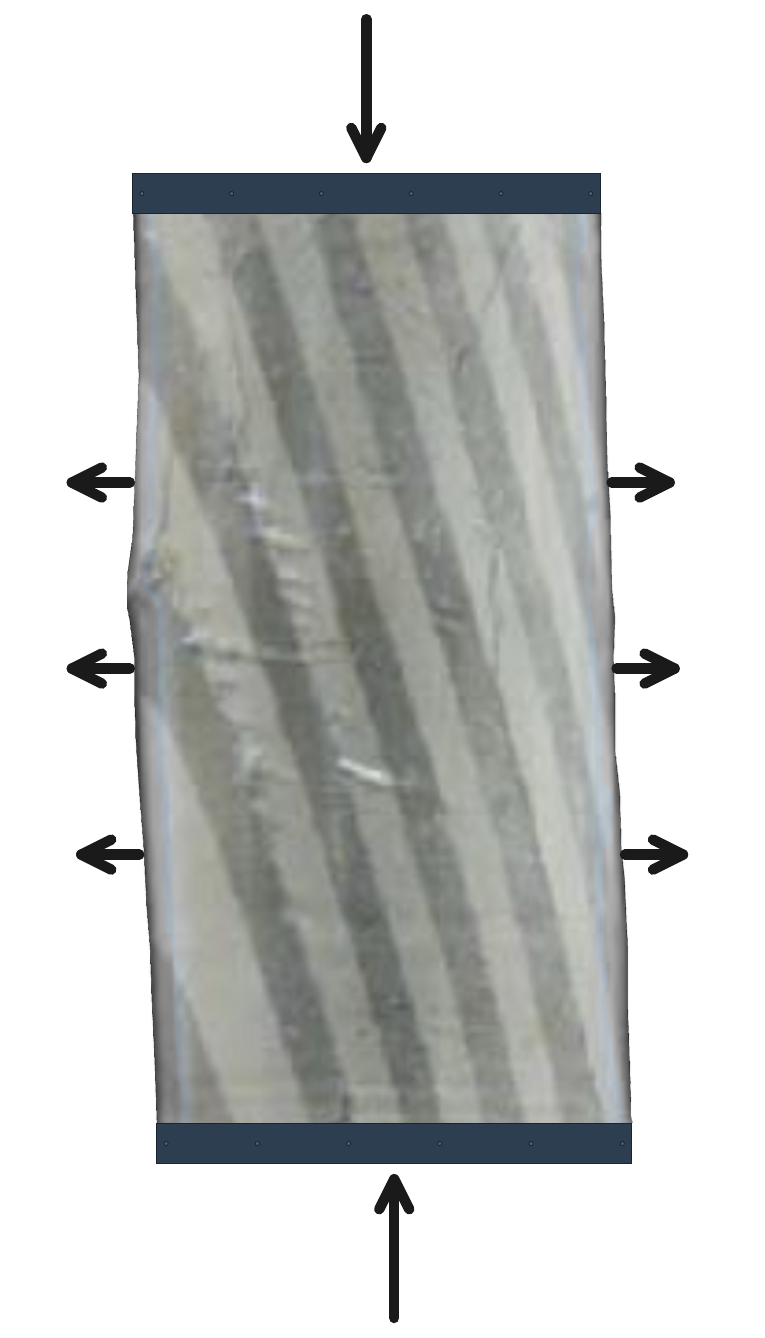}
        \caption{}
        \label{fig:schematic_crushing}
    \end{subfigure}
    \caption{The wrinkling / splitting analogy: (a) global tension with constrained lateral contraction induces local transverse compression and wrinkling; (b) conversely, global compression with constrained lateral expansion induces local transverse tension and axial splitting (figure from \cite{Tien2006}).}
    \label{fig:wrinkling_inverse_schematic}
\end{figure}

Decades of research have established three principal theoretical frameworks to explain compressive fracture, each with explanatory power but distinct limitations. The dominant micromechanical paradigm focuses on flaw-initiated fracture, recognizing that brittle failure occurs well below the ideal strength because defects localize stress in the sense of Griffith~\cite{griffith1921phenomena}. The canonical ``wing-crack'' model posits that sliding on inclined microcracks nucleates tensile cracks that grow parallel to the principal compression~\cite{Hoek1965,Nemat-Nasser1982} . Detailed analyses show these cracks curve axially, driving splitting under low confinement, while lateral compression arrests them~\cite{horii1985compression,renshaw2001universal}, and elastic heterogeneity can drive similar in-plane shear propagation even without friction~\cite{janach1980plane}. This framework has been foundational for decades~\cite{Ashby1986, Wang2004} and remains integral to modern constitutive models that incorporate rate-dependence~\cite{Bhat2012JAM,Rosakis2000SST} or plasticity~\cite{Deshpande2008Ceramics}. However, its underlying mechanics are fundamentally stochastic. It relies on unknown flaw distributions~\cite{tang2000influence} and statistical ``avalanche'' mechanics~\cite{Sethna2001Nature}. Crucially, it struggles to explain early-stage axial cracking~\cite{Brace1966}, and its driving mechanism can rapidly arrest, producing stable cracks rather than macroscopic failure~\cite{brace1963note, lehner2001winged}. This focus on micro-scale singularities overlooks the deterministic role of continuum-level stress fields, a limitation similarly observed in metamaterial fracture~\cite{Shaikeea2022NatMat}.

A second major paradigm attempts to explain macroscopic splitting through global energetic criteria; however, these approaches fundamentally lack a mechanistic trigger for failure. These models frame compressive splitting as a thermodynamic competition between intact and split configurations~\cite{mikhalyuk1997dissipation, huang2014conversion, wasantha2014energy}. For instance, a heterogeneous solid is predicted to split when the release of stored microstructural elastic energy exceeds the cost of new fracture surfaces~\cite{Bhattacharya1998}. While this framework possesses strong predictive power and provides a compelling thermodynamic rationale, it does not specify a kinematic pathway. It identifies the energetic driving force for failure, but not the mechanical driver that initiates it.

Finally, classical criteria rooted in rock and soil mechanics, such as Mohr-Coulomb~\cite{Hoek1965}, assume failure occurs via shear rupture when inclined stresses overcome intrinsic cohesion and friction. While effective under confinement and for explaining angled shear faults, this purely shear-based framework cannot explain axial splitting. It fundamentally contradicts experimental observations in brittle materials like granite, which develop open, axial tensile cracks before ultimate failure~\cite{Brace1966}; moreover, as a pointwise criterion it carries no dependence on specimen geometry.

These prior approaches highlight an important theoretical gap: flaw-based models are inherently stochastic, while energetic and shear-based models lack a specific mechanical trigger for splitting. Seeking to overcome the limitations of stochastic failure models, emerging work demonstrates that deterministic geometric design dictates mechanical response. This shift has yielded phenomena such as ``microbuckling'' of granular crystals~\cite{Karuriya2022PNAS}, ``quasi-ductile'' fracture via jigsaw interfaces in glass~\cite{Mirkhalaf2014PNAS}, and ``tablet-sliding'' in nacre-like composites~\cite{Yin2019Science}.

However, even in nominally homogeneous bulk solids, macroscopic boundary conditions deterministically precipitate failure. In standard compression tests, loading platens invariably constrain the lateral Poisson expansion, inducing a non-uniform stress state characterized by significant transverse tension~\cite{Swab2025}; indeed, experimentally eliminating this end friction alters both the lateral stress state and fracture regularity~\cite{fakhimi2015axial}. The fact that this macroscopic, continuum-level effect dominates the failure response motivates our search for a deterministic trigger that does not rely on pre-existing micro-flaws. 

In this Letter, we propose that boundary-induced tensile stresses act as the deterministic trigger for axial splitting. 
Our central hypothesis is that global compression, when kinematically constrained, induces a localized transverse tension. 
This mechanism is the compressive inverse of the well-established wrinkling instability in thin elastic sheets~\cite{friedl2000buckling, Cerda2002Nature, Li2011PRL, Audoly2010Book}. 
Just as clamping a stretched sheet prevents lateral Poisson contraction and consequently compressive buckling, constrained compression prevents lateral Poisson expansion and consequently a localized transverse tension that drives Mode I (opening) fracture (Fig.~\ref{fig:wrinkling_inverse_schematic}).
That is, global tension with constrained contraction leads to local compression and wrinkling, whereas global compression with constrained expansion generates transverse tension and splitting.
By identifying a deterministic stress feature as the mechanical trigger, this framework supplies the missing kinematics for global energy models, unifying the energetic driving force with a specific physical mechanism.

This framework challenges the foundational assumptions of classical invariant-based failure criteria. Pressure-sensitive yield criteria such as Drucker--Prager~\cite{Drucker1952,Prager1949} incorporate mean stress effects but do not describe brittle fracture or compressive splitting. Furthermore, the Christensen failure theory~\cite{Christensen2013Book}, which provides a general envelope for homogeneous materials, constructs its polynomial-invariant surface by prescribing tensile ($T$) and compressive ($C$) strengths as strictly independent inputs. Our results demonstrate that this assumed independence is not fundamental for brittle solids under confined compression. Under the boundary conditions of standard compression tests, the apparent compressive strength arises directly from the boundary-induced tensile stress; the global compressive load naturally generates the transverse tensile field that activates Mode I splitting. Consequently, in contrast to classical criteria where tension--compression asymmetry is imposed phenomenologically, our framework predicts a fundamental mechanical relationship between the two strengths. The compressive splitting strength emerges deterministically from continuum elasticity, specimen geometry, and the Poisson effect, rather than as an independent material parameter.

%%%%%%%Methodology %%%%%%%%%%%%%
%%%%%%%%%%%%%%%%%%%%%%%%%%%%%%%
To formalize this mechanism, the problem is idealized as a homogeneous, isotropic elastic block compressed in the vertical $y$-direction between two rigid platens. 
These boundaries are ``clamped,'' meaning interfacial friction completely prevents the lateral displacement of the top and bottom surfaces ($y = \pm H/2$). 
The pre-fracture stress state is determined by solving this linear elastic boundary value problem using a superposition principle, directly inverting the classical wrinkling approach~\cite{friedl2000buckling}.

First, consider the unconstrained state subjected to a uniform uniaxial compressive stress $\sigma_y = -\sigma_c$.
The block would undergo a uniform lateral expansion in the transverse ($x$) direction due to the Poisson effect. 
This unconstrained displacement, $u_x^{\text{free}}$, is given by:
\begin{equation}
    u_x^{\text{free}}(x) = -\nu \epsilon_y x = \nu \frac{\sigma_c}{E} x 
\end{equation}
which is maximum at the free edges $x = \pm B/2$.

Next, we impose the clamped boundary conditions. To prevent lateral expansion at $y = \pm H/2$ (the displacement-constrained platens), a ``restoring'' stress field must be superimposed to counteract $u_x^{\text{free}}$. 
Solving the equations of linear elasticity for this restoring stress reveals a complex stress distribution. 
Crucially, a region of lateral \textit{tensile} stress ($\sigma_{xx} > 0$) develops away from the clamped ends, concentrating in the interior. 

\begin{figure}[ht!]
    \centering
    \subfloat[]{\label{fig:stress-yy}\includegraphics[width=0.45\columnwidth]{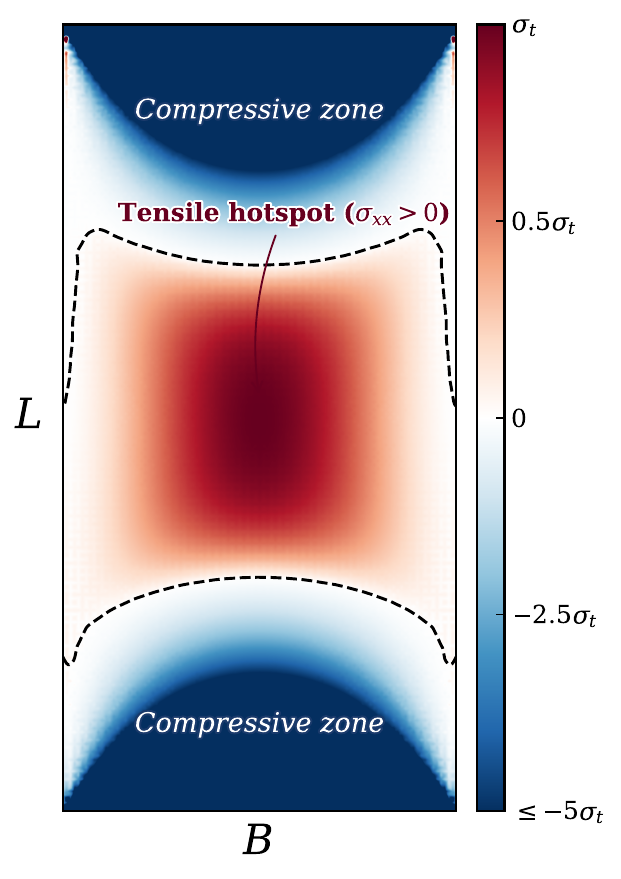}}
    \hfill
    \subfloat[]{\label{fig:stress-xy}\includegraphics[width=0.45\columnwidth]{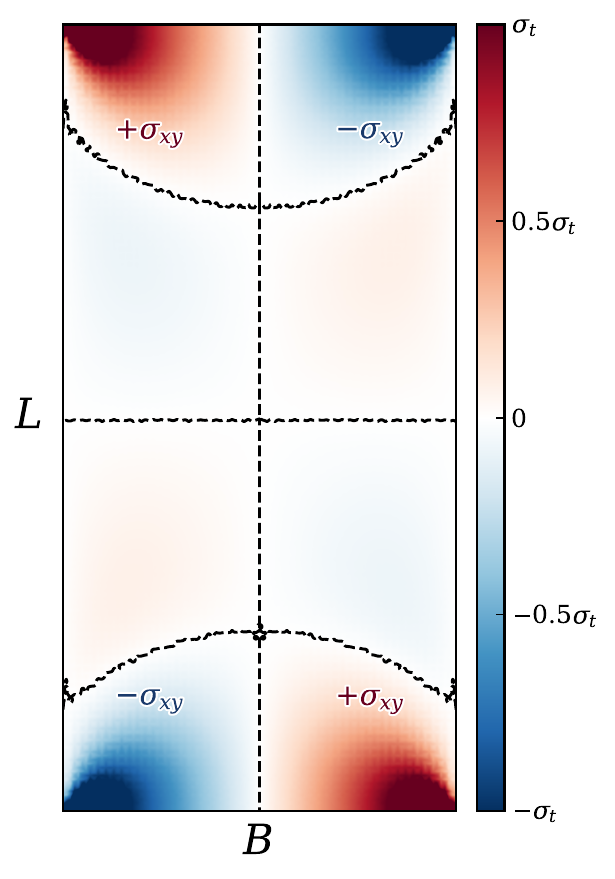}}
    \caption{Stress distribution using finite element analysis for an aspect ratio $L/B=2$ and $\nu = 0.3$, compressed along the $y$-axis with clamped boundaries. (a) The normalized transverse stress ($\sigma_{xx} / \sigma_t$) reveals an interior region under tension (positive values, red), with the dashed line indicating the zero-stress contour. (b) The corresponding normalized shear stress ($\sigma_{xy} / \sigma_t$).}
    \label{fig:fem_results}
\end{figure}

\begin{figure}[ht!]
    \centering
    \includegraphics[width=0.82\columnwidth]{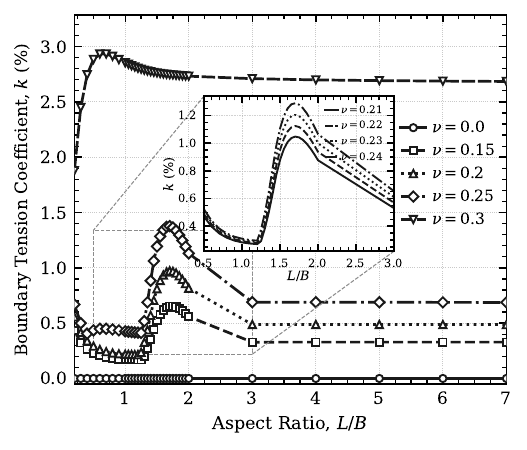} 
    \caption{The effect of specimen aspect ratio ($L/B$) and Poisson's ratio ($\nu$) on the maximum induced transverse tension, $k$. A peak occurs at intermediate aspect ratios.}
    \label{fig:k_vs_aspect_ratio}
\end{figure}

Using finite element analysis (Fig.~\ref{fig:fem_results}) for a specimen with an aspect ratio of $L/B=2$ and $\nu = 0.3$, we find that this constrained compression generates a central tensile ``hotspot''. 
This tension under axial compression is analogous to that arising in the Brazilian disk test, where diametrically opposed loads generate transverse tension along the loading diameter~\cite{Johnson1985}; the present mechanism is distinctly Poisson-driven, however, as evidenced by the vanishing of the boundary-induced tension at $\nu = 0$ (Fig.~\ref{fig:k_vs_aspect_ratio}). 
This localized transverse tension provides the mechanical pathway to initiate Mode I axial splitting.

To quantify the boundary-induced tension, we define a dimensionless coefficient, $k=\sigma_{x x, \max } / \sigma_c$, which quantifies the ``efficiency'' with which global compression is converted into localized transverse tension; by dimensional analysis of the linear elastic problem, $k$ depends only on geometry ($L/B$) and Poisson's ratio ($\nu$), and is independent of the elastic modulus $E$. This is evaluated via plane-strain finite element simulations, appropriate for the solid, non-thin specimens used in the experimental compression tests being modeled.
As shown in Fig.~\ref{fig:k_vs_aspect_ratio}, the dependence of $k$ on the specimen aspect ratio ($L/B$) and Poisson's ratio ($\nu$) reveals a non-monotonic and strongly non-separable coupling, exhibiting qualitative mode transitions analogous to those recently discovered in tensile buckling~\cite{rammerstorfer2024buckling}. 
For low $\nu$, $k$ peaks at $L/B \approx 1.7$, corresponding to a long-wavelength ``boundary interaction'' mode. 
As $\nu$ increases, the enhanced lateral expansion generates significantly larger restoring stresses, and this peak shifts to much lower aspect ratios ($L/B \lesssim 1$), indicating a transition to a ``local Poisson'' mode dominant in squat geometries.
In all cases, $k$ asymptotically converges for slender specimens ($L/B \ge 3$), consistent with Saint-Venant's principle.

\begin{figure}[ht!]
    \centering
    \includegraphics[width=0.95\columnwidth]{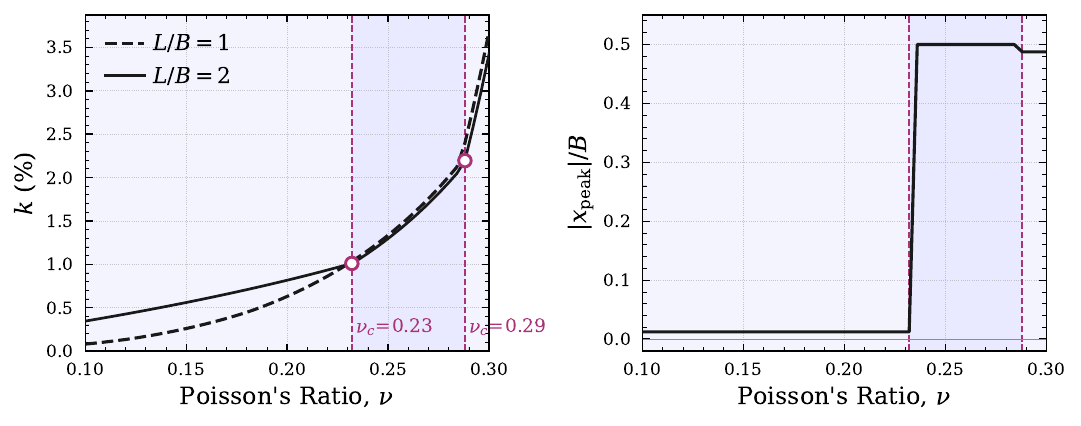} 
    \caption{Mode transition analysis for specimens with $L/B=1.0$ and $L/B=2.0$. (left) The coefficient $k$ versus $\nu$. (right) The location of the peak stress, $|x_{peak}|/B$ for $L/B=2.0$ case. Dash lines correspond to two kinks in the $k$ plot for $L/B=2.0$.}
    \label{fig:bifurcation_analysis}
\end{figure}

This shift signifies a complex mode competition within the experimentally relevant range ($1 \le L/B \le 2$). Plotting $k$ versus $\nu$ (Fig.~\ref{fig:bifurcation_analysis}, left) reveals a distinct crossover at $\nu \approx 0.23$, where the optimal geometry for tension generation switches from $L/B=2$ to $L/B=1$: below this threshold, the slender geometry  ($L/B = 2$) generates larger boundary-induced tension; above it, the squat geometry ($L/B = 1$) becomes more efficient. The crossover at $\nu\approx 0.23$ is not universal but depends on the specific pair of geometries compared -- a consequence of the continuous migration of the $k$ peak toward smaller $L/B$ as $\nu$ increases, as illustrated by the additional curves near the transition in  Fig.~\ref{fig:k_vs_aspect_ratio} (inset).

Furthermore, the $k$-curves exhibit kinks corresponding to sharp bifurcations in the pre-fracture stress field.
By tracking the transverse location of the peak tensile stress, $x_{peak}$ (Fig.~\ref{fig:bifurcation_analysis}, right), we observe that the tension maximum remains geometrically centered ($x_{peak}=0$) until a critical $\nu \approx 0.23$, beyond which it undergoes a discontinuous symmetry-breaking transition to an off-center position; $|x_{peak}|/B$ is plotted as the natural order parameter of this bifurcation, collapsing the two physically equivalent solutions $x_{peak} = \pm\delta$ onto a single branch.

This symmetry-breaking bifurcation in the spatial topology of the boundary-induced stress field -- and the cascade of further transitions at higher $\nu$ -- mirrors the geometry-sensitive crossovers between necking and wrinkling modes identified in tensile elastic solids~\cite{rammerstorfer2024buckling, Riccobelli2024}, here emerging entirely within the linear elastic pre-fracture regime of a brittle solid under compression.

\begin{figure}[ht!]
    \centering
    \includegraphics[width=0.76\columnwidth]{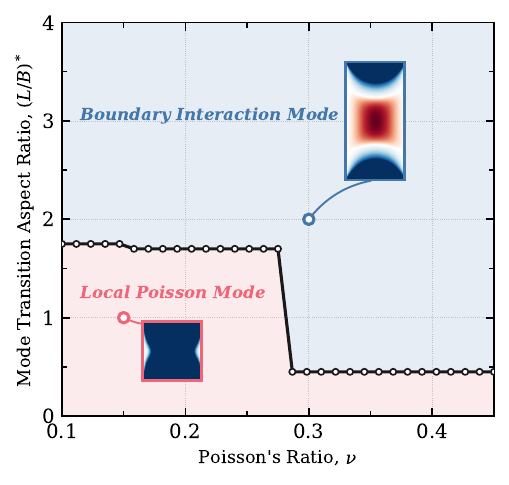}
    \caption{Mode map in the $(\nu,\, L/B)$ parameter space. The curve marks the aspect ratio $(L/B)^*$ at which the boundary-induced tension coefficient $k$ is maximized, separating the \textit{boundary interaction} (above) and \textit{local Poisson} (below) regimes. Insets show the $\sigma_{xx}$ field for representative specimens in each regime.}
    \label{fig:mode_map}
\end{figure}

Visualizing this in parameter space (Fig.~\ref{fig:mode_map}) yields a mode map separated into the low-$\nu$ boundary interaction regime and the high-$\nu$ local Poisson regime, a direct compressive analog to necking-wrinkling crossovers in stretched elastic sheets~\cite{Riccobelli2024}. Near $\nu \approx 0.28$, a secondary local maximum of $k(L/B)$ corresponding to the local Poisson mode near $L/B \approx 0.5$ overtakes the boundary interaction peak near $L/B \approx 1.7$, causing $(L/B)^*$ to drop discontinuously to values near $0.5$. Since standard compression test specimens satisfy $L/B \geq 1$, they are on the descending branch of $k(L/B)$, generating less boundary-induced tension than the geometric optimum and thereby suppressing axial splitting relative to a squat specimen; further details are provided in the Supplemental Material~\cite{SupplementalMaterial}

In practical geological and engineering settings, lateral confining pressure ($\sigma_{\text{conf}}$) introduces a competing mechanism. 
Under superposition -- which is exact in linear elasticity for fixed geometry and material -- the transverse stress state is the sum of internally induced Poisson tension and externally applied transverse compression. 
The critical confinement ratio at which tension is fully suppressed follows directly as $\sigma_{\text{conf}}/\sigma_c = 
k(L/B,\nu)$, which for $L/B=2.0$ and $\nu=0.3$ gives approximately $2.0\%$. Beyond this ratio, the specimen is in a state of pure compression and tensile splitting is entirely suppressed. 
This provides a rigorous continuum mechanics explanation for the well-documented experimental observation that lateral confinement significantly enhances the compressive strength and ductility of brittle materials by entirely inhibiting the primary tensile splitting mode. Consistently, triaxial extension experiments on rock document a continuous transition from axial splitting to shear fracture with increasing confining pressure~\cite{RamseyChester2004, Zeng2019}, precisely as the boundary-induced tension is progressively suppressed by the applied lateral compression.

Returning to the low-confinement setting (Fig.~\ref{fig:fem_results}), the maximum transverse tension is precisely at the geometric center. 
Assuming a maximum principal stress criterion for brittle fracture, failure initiates at this central point. This localized tension drives a Mode I fracture, resulting in a vertical fracture plane oriented parallel to the loading axis. This provides a direct, continuum-based explanation for axial splitting that aligns with experimental observations of internal crack nucleation~\cite{Brace1966}. Furthermore, this tensile-driven splitting mechanism is fundamentally distinct from the shear-driven formation of conical faults often observed under high confinement, which typically initiate via stress concentrations at the corners~\cite{Chen1996}.

We establish a predictive failure criterion by connecting this pre-fracture elasticity solution 
to the material's intrinsic tensile strength, $\sigma_t$. Fracture nucleates when the 
boundary-induced tension reaches the intrinsic tensile threshold---a deterministic, 
geometry-controlled condition that replaces stochastic flaw-based nucleation criteria. 
The clamped boundaries generate a localized transverse tension that is an inevitable 
geometric consequence of the kinematic constraint, providing the specific mechanical 
trigger that global energy models lack. Specifically, fracture nucleates when this 
boundary-induced tension, $\sigma_{{xx}, \text{max}}$, reaches the intrinsic threshold:
\begin{equation}
    \sigma_{\text{xx,max}} = \sigma_t
    \label{eq:predictive_criteria}
\end{equation}
This addresses nucleation from a nominally intact solid, and is distinct from the energetic Griffith model for the growth of existing cracks ~\cite{griffith1921phenomena,binetti2025thermal}; prior work has shown that crack initiation is not purely energetic \cite{salman2021delocalizing, puglisi2013cohesion, del2013variational,leguillon2002strength, weissgraeber2016review, 
doitrand2022dynamic, dormieux2025remarks}.
To obtain a transparent relation that exposes the key physics, we treat the tensile stress as a simple proxy for the initiation of failure~\cite{miehe2010thermodynamically, DeLorenzis2022, Steinke2019, hakimzadeh2022phase,hakimzadeh2025crack,hakimzadeh2025phase}.
Further, we notice by symmetry that $\sigma_{xy} = 0$ where the largest tensile stress develops (Fig.~\ref{fig:fem_results}b), so  Eq.~\eqref{eq:predictive_criteria} follows directly.

Using superposition, the total maximum transverse tension, incorporating both the boundary-induced tension and the external confining pressure ($\sigma_{\text{conf}}$), is:
\begin{equation}
    \sigma_{\text{xx,max}} = k(L/B, \nu) \cdot \sigma_c - \sigma_{\text{conf}}
\end{equation}

As established in our parametric study, $k$ exhibits a complex, non-separable coupling. Inspired by analytical treatments of buckling instabilities~\cite{friedl2000buckling}, we investigated whether this could be simplified by decomposition into a separable form, $k = f(\nu) \cdot h(L/B)$; however, various physically-motivated scaling laws yielded no successful data collapse.  This  non-separability is a direct structural consequence of the symmetry-breaking mode transition 
established above: because $\nu$ fundamentally alters the geometric mode of tension generation -- shifting the peak location in $L/B$ rather than simply scaling its 
magnitude -- no factored form can exist by construction. 
The Poisson's ratio does not simply scale the magnitude of the stress, but fundamentally alters its geometric dependence, meaning a simple expression for $k$ does not exist.

By combining the failure criterion with the superposition model at the point of failure (where $\sigma_c = \sigma_{c,crit}$), we solve for the critical compressive splitting strength:
\begin{equation}
    \sigma_{c,crit} = \frac{1}{k(L/B, \nu)} (\sigma_t + \sigma_{\text{conf}})
    \label{eq:predictive_model}
\end{equation}
Eq.~\ref{eq:predictive_model} predicts the onset of the macroscopic splitting instability from the nominally intact solid, as distinct from the stable, distributed microcracking documented well below peak load~\cite{Brace1966}. For the brittle solids considered here, ultimate failure
($\sigma_1$ in experiments) follows this onset closely.
This establishes a direct relationship between axial splitting strength and confining pressure, and demonstrates that compressive strength is not an independent material parameter but can be derived entirely from the tensile strength, $\nu$, and the specimen geometry.

A particularly transparent consequence of Eq.~\eqref{eq:predictive_model} is obtained in the unconfined case ($\sigma_{\text{conf}} = 0$). The resulting macroscopic compressive splitting strength is $C \equiv \sigma_{c} = \sigma_t / k(L/B,\nu)$, which strictly fixes the strength ratio:
\begin{equation}
    \frac{C}{T} = \frac{1}{k(L/B,\nu)}
    \label{eq:C_over_T_from_k}
\end{equation}

For axial compression with clamped end constraints, Eq.~\eqref{eq:C_over_T_from_k} shows that the commonly used strength ratio $C/T$ is not an independently assigned material property. This has profound implications for invariant-based failure criteria, such as Drucker--Prager~\cite{Drucker1952, Prager1949} and Christensen~\cite{Christensen2013Book}, which treat uniaxial tensile and compressive strengths $T$ and $C$ as independent inputs when constructing a failure surface. In the present setting, our relation provides a mechanical closure relation. When used to eliminate $C$ in favor of $T$ and $k(L/B,\nu)$, the traditional two-parameter phenomenological strength description collapses into a one-parameter family controlled solely by $T$ and the specimen configuration. This geometry-dependence of measured compressive strength is independently documented 
in standardized testing~\cite{ASTMC39}, where empirical aspect-ratio correction factors have been prescribed without a mechanics basis; Eq.~\eqref{eq:C_over_T_from_k} now supplies that basis.

To test our hypothesis, we compare against experimental data for a diverse set of brittle materials: a synthetic rock~\cite{Tien2006}, sintered aluminum nitride~\cite{Chen1996}, rocksalt~\cite{Handin1953}, a glass-ceramic~\cite{Lankford1995, Corning1971}, silicon carbide~\cite{Lee2005}, and Westerly granite~\cite{Brace1966}. Because our continuum framework models an ideal, flawless solid, it naturally establishes an upper bound on strength. To isolate the model's predicted physical trend from the absolute flaw-limited strength of real materials, we introduce a single best-fit strength reduction factor, $\alpha$, determined via linear regression for each material ($\sigma_{c,crit}^{\prime}=\alpha \cdot\sigma_{c,crit}$).

\begin{figure}[ht!]
    \centering
    \includegraphics[width=0.95\columnwidth]{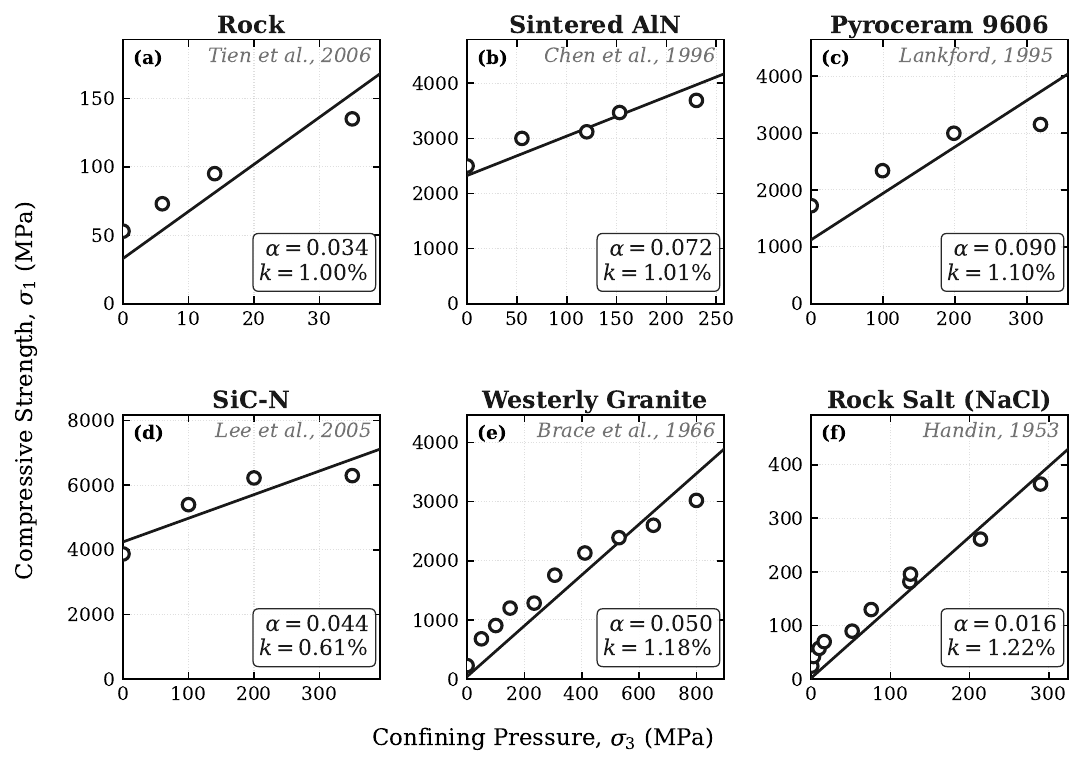}
    \caption{Comparison of the model's predicted compressive strength (solid lines) against experimental data (markers) for a range of brittle materials.}
    \label{fig:model_vs_experiment_scaled_regression}
\end{figure}

The results (Fig.~\ref{fig:model_vs_experiment_scaled_regression}) demonstrate strong alignment between the rescaled predictions and the experimental data. This validates our continuum mechanism, showing that it  captures the fundamental physics of how confinement enhances compressive strength. Notably, $\alpha$ rescales absolute strength but preserves the predicted functional form: the ratio of confinement sensitivity 
to zero-confinement strength is fixed at $1/\sigma_t$, independent of fitting. Across all six materials $\alpha \sim O(10^{-2})$, confirming that the continuum mechanism  captures the confinement scaling while $\alpha$ absorbs flaw-limited strength reduction, boundary condition imperfections common to real compression tests, and the short flaw-dependent interval between splitting onset and peak load.
The Supplemental Material~\cite{SupplementalMaterial} discusses auxetic materials ($\nu < 0$) as well as two structural advantages of our framework over classical energetic models~\cite{Bhattacharya1998}: our strength--confinement relation is parameterized solely by measurable continuum properties ($E$, $\nu$, $\sigma_t$, $L/B$), whereas the energetic model additionally requires the elastic moduli of the notionally cracked phase and the ratio between surface energy and microstructure scale; and our model remains valid even in the high-confinement regime where the energetic criterion is not.

%%%%%%%%%%%%%%%%%%%%%
%%%%%%%%%%%%%%%%%%%%%
%%%%%%%%%%%%%%%%%%%%%
The fracture morphology reported in ~\cite{Heard1980} provides direct validation of this deterministic mechanism. At low confinement, BeO  failed by tensile fracture parallel to the loading axis; however, at higher confinement (0.4--0.7 GPa), the specimens exhibited distributed slip with no observable axial cracks. These observations correspond precisely to our predicted stress fields, where the central band of transverse tension ($\sigma_{xx}>0$) disappears as confinement suppresses the boundary-induced tension. The experimental transition from axial splitting to crushing therefore aligns directly with the suppression of the macroscopic tensile hot spot. 

Crucially, this boundary-induced mechanism offers a deterministic continuum alternative to classical statistical fracture theories. Stochastic ``weakest-link'' models (e.g., Weibull statistics) are notoriously unstable in a renormalization-group sense for quasibrittle materials because they neglect stress-based flaw interactions~\cite{Bertalan2014PRApp}. Furthermore, modern statistical physics reveals that diffuse, ``avalanche''-based damage is a finite-size effect; in the thermodynamic limit, failure reverts to a nucleation-driven process~\cite{Shekhawat2013PRL}. Our framework supplies the initiator for this continuum-scale nucleation in engineering materials with relatively narrow flaw distributions (Weibull modulus $>3$) ~\cite{meyers2008mechanical, lawn1993fracture}: a deterministic, macroscopic tensile stress feature ($\sigma_{xx,max}$) that is a purely geometric consequence of the boundary constraints.

This deterministic, geometry-controlled picture parallels the principle of ``geometry-over-flaws'' in lattice and architected materials~\cite{Schaedler2011Science}, where failure is governed by a geometric length scale, with material flaws secondary. 
In those systems, the length scale is set by the architecture that is typically coarse enough that scale separation does not hold~\cite{Shaikeea2024JMPS, Mirkhalaf2014PNAS, Yin2019Science, Zhu2024Bioinspir, Karuriya2022PNAS}. 
The materials considered here \textit{do} satisfy scale separation (see Supplemental Material~\cite{SupplementalMaterial}), nonetheless their compressive strength is strongly geometry-controlled and driven by the loading boundaries.

%%%%%%%%%%%%%%%%%%%%%
%%%%%%%%%%%%%%%%%%%%%
%%%%%%%%%%%%%%%%%%%%%
%%%%%%%%%%%%%%%%%%%%%
%%%%%%%%%%%%%%%%%%%%%
%%%%%%%%%%%%%%%%%%%%%
%%%%%%%%%%%%%%%%%%%%%
%%%%%%%%%%%%%%%%%%%%%

\begin{acknowledgments}
    We acknowledge ARO (MURI W911NF-24-2-0184) for financial support; G. Ravichandran and Tony Rollett for insightful discussions; and NSF ACCESS (MCH240078) for computing resources provided by the Pittsburgh Supercomputing Center.   
\end{acknowledgments}

\bibliographystyle{unsrt}
\bibliography{crushing}
%%%%%%%%%%%%%%%%%%%%%
%%%%%%%%%%%%%%%%%%%%%
%%%%%%%%%%%%%%%%%%%%%
%%%%%%%%%%%%%%%%%%%%%
\appendix

\makeatletter
\renewcommand*{\thesection}{\Alph{section}}
\renewcommand*{\thesubsection}{\thesection.\arabic{subsection}}
\renewcommand*{\p@subsection}{}
\renewcommand*{\thesubsubsection}{\thesubsection.\arabic{subsubsection}}
\renewcommand*{\p@subsubsection}{}
\makeatother

\clearpage
\onecolumngrid

\section{Supplementary Material}

\subsection*{Scaling Analysis of $k(L/B, \nu)$}

In the main text, we assert that the boundary tension coefficient $k(L/B, \nu)$ cannot 
be simplified into a separable form $f(\nu) \cdot h(L/B)$. Here, we present the evidence 
for this conclusion by attempting to collapse the finite element data onto a single master 
curve using three physically motivated scaling laws derived from continuum mechanics.

If the geometric and material effects were decoupled, the normalized coefficient 
$k_\mathrm{scaled}$ would be independent of $\nu$. We tested three scaling hypotheses:

\begin{enumerate}
    \item \textbf{Linear Poisson Scaling:} $k_\mathrm{scaled} = k / \nu$. This assumes 
    the boundary-induced tension scales linearly with the lateral expansion drive.

    \item \textbf{Plane Strain Stiffness Scaling:} $k_\mathrm{scaled} = k \cdot (1-\nu^2)$. 
    This normalizes the boundary-induced tension by the effective plane-strain compliance, 
    testing whether the stress concentration magnitude is governed by the 2D stiffness.

    \item \textbf{Volumetric Constraint Scaling:} $k_\mathrm{scaled} = k \cdot 
    \frac{(1+\nu)(1-2\nu)}{\nu}$. This scaling is derived from the Lam\'{e} parameters 
    governing volumetric strain, testing whether the boundary-induced tension is driven 
    by incompressibility effects (divergence as $\nu \to 0.5$).
\end{enumerate}

\vspace{0.56em}
\noindent
We note that $(1-\nu^2)$ factors also arise in classical plane-strain fracture energetics 
(e.g., $\Gamma_c = (1-\nu^2)K_I^2/E$ for mode-I loading), underscoring that constraint 
effects can make $\nu$ enter in nontrivial, non-separable ways~\cite{becker2025tectonic}.

\begin{figure}[htbp]
    \centering
    \includegraphics[width=\textwidth]{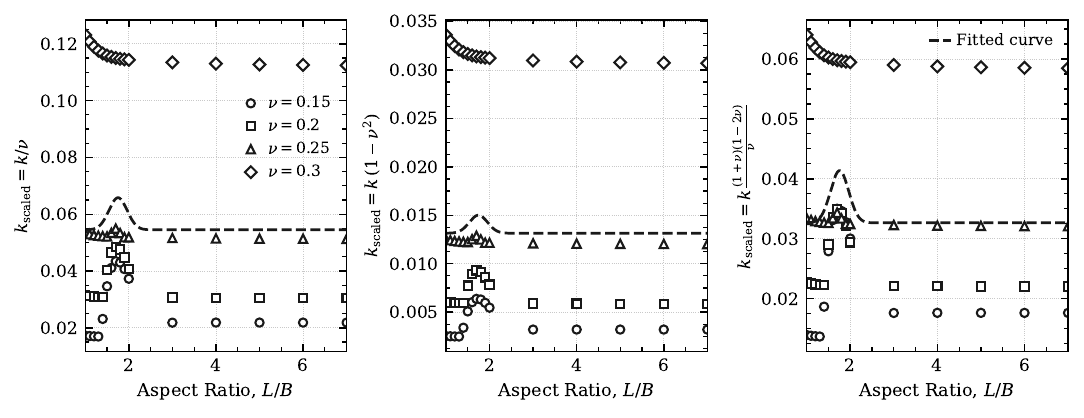}
    \caption{Attempts to collapse the boundary tension coefficient $k$ onto a 
    single master curve using three scaling laws motivated by continuum mechanics. 
    The dashed line shows the best-fit Gaussian$+$constant to the mean of the 
    scaled data. The failure of the scaled data to align — particularly the shift 
    in the peak location along the $L/B$ axis — confirms that material properties 
    and geometry are non-separably coupled.}
    \label{fig:supp_scaling_attempts}
\end{figure}

Figure~\ref{fig:supp_scaling_attempts} presents the results of these scaling attempts. 
While the asymptotic behavior for long specimens ($L/B > 3$) collapses reasonably well 
under linear scaling (Fig.~\ref{fig:supp_scaling_attempts}), the region $L/B < 2$ fails to 
collapse in all cases. Notably, the aspect ratio at which $k$ is maximized shifts 
significantly with $\nu$:

\begin{itemize}
    \item For low $\nu$, the peak occurs near $L/B \approx 1.7$.
    \item For high $\nu$, the peak shifts to $L/B \lesssim 1.0$.
\end{itemize}

Because the location of the peak along the $L/B$ axis changes with $\nu$, no $y$-axis 
rescaling $f(\nu)$ can align the curves. This is a direct structural consequence of the 
symmetry-breaking mode transition established in the main text: because $\nu$ 
fundamentally alters the \emph{geometric mode} of the boundary-induced tension 
concentration — shifting the peak location in $L/B$ rather than simply scaling its 
magnitude — no factored form $f(\nu)\cdot h(L/B)$ can exist by construction. The 
non-separability therefore reflects the coupled dependence of the boundary-induced stress 
concentration on both material and geometry across the full parameter space.

\clearpage

\subsection*{Mode Structure of the Boundary Tension Coefficient}

The main text identifies two qualitative regimes in the 
dependence of $k$ on $(L/B, \nu)$: a boundary interaction 
mode dominant at low $\nu$ and a local Poisson mode dominant at high $\nu$. Here we provide a more detailed characterization of this mode structure and its consequences.

\begin{figure}[htbp]
    \centering
    \includegraphics[width=0.6\linewidth]{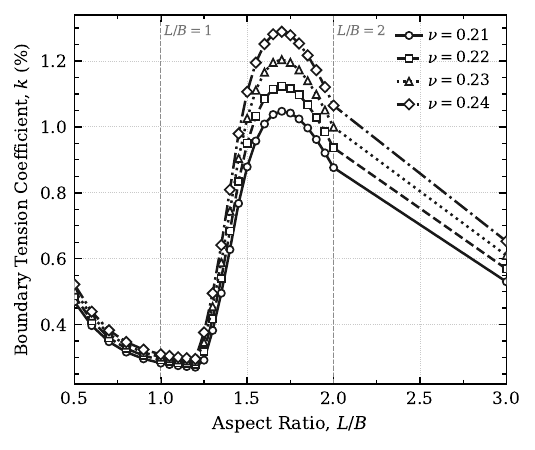}
    \caption{Boundary tension coefficient $k$ versus 
    $L/B$ for $\nu \in \{0.21, 0.22, 0.23, 0.24\}$, 
    zoomed into $L/B \in [0.5, 3.0]$. Dashed vertical 
    lines mark the two standard aspect ratios $L/B = 1$ 
    and $L/B = 2$. The peak migrates continuously toward 
    lower $L/B$ as $\nu$ increases, confirming the mode 
    transition is a smooth process.}
    \label{fig:supp_inset_detail}
\end{figure}

\paragraph*{Continuous nature of the mode transition.}
Figure~\ref{fig:supp_inset_detail} shows $k$ versus $L/B$ 
for $\nu \in \{0.21, 0.22, 0.23, 0.24\}$, zooming into 
the transition region $L/B \in [0.5, 3.0]$ with reference 
lines at the two standard aspect ratios $L/B = 1$ and 
$L/B = 2$. The peak of $k$ shifts continuously toward 
lower $L/B$ as $\nu$ increases --- there is no 
discontinuous jump in the peak location within this range. 
The crossover in Fig.~\ref{fig:bifurcation_analysis} 
(left panel), where the $L/B = 1$ curve overtakes the 
$L/B = 2$ curve at $\nu \approx 0.23$, therefore reflects 
a smooth migration of the optimal geometry rather than an 
abrupt transition. The crossover value $\nu \approx 0.23$ 
is not a universal material constant: it is the value at 
which $k(L/B=1,\nu) = k(L/B=2,\nu)$ for this specific 
geometry pair. The inset of Fig.~\ref{fig:k_vs_aspect_ratio} 
in the main text confirms that different aspect ratio 
comparisons yield different crossover values.

\begin{figure}[htbp]
    \centering
    \includegraphics[width=\linewidth]{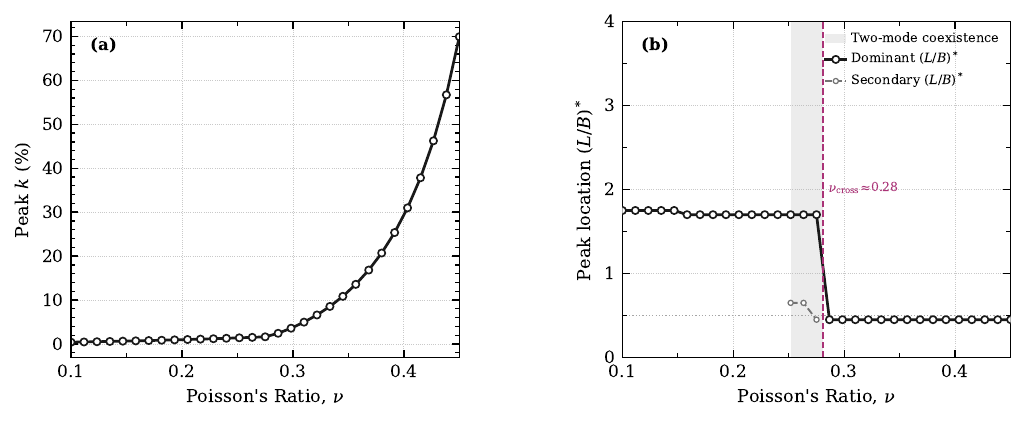}
    \caption{Extended parametric analysis across 
    $\nu \in [0.10, 0.45]$.
    (\textit{a})~Peak $k$ versus $\nu$; rapid growth 
    as $\nu \to 0.5$ reflects divergence of the 
    plane-strain bulk modulus $E/(2(1-2\nu))$.
    (\textit{b})~Mode transition aspect ratio $(L/B)^*$ 
    versus $\nu$. Grey band: two-mode coexistence 
    $\nu \in [0.25, 0.28]$; dashed grey line: secondary 
    (local Poisson) peak location; purple dashed line: 
    crossover $\nu_\mathrm{cross} \approx 0.28$. 
    For $\nu \gtrsim 0.28$, $k(L/B)$ varies by less 
    than 5\% across $L/B \approx 0.45$--$0.90$, so 
    $(L/B)^*$ represents an approximate centroid of a 
    broad flat optimum.}
    \label{fig:supp_mode_map_extended}
\end{figure}

\paragraph*{Evolution of $(L/B)^*$ and mode competition.}
Figure~\ref{fig:supp_mode_map_extended} shows $(L/B)^*$ --- 
the aspect ratio at which $k$ is maximized for a given 
$\nu$ --- across $\nu \in [0.10, 0.45]$ 
(Fig.~\ref{fig:supp_mode_map_extended}b), together with the 
peak value of $k$ (Fig.~\ref{fig:supp_mode_map_extended}a).

For $\nu \lesssim 0.25$, $k(L/B)$ has a single dominant 
maximum near $L/B \approx 1.7$ (boundary interaction mode), 
consistent with Fig.~\ref{fig:mode_map} of the main text.

For $\nu \in [0.25, 0.28]$, finite-element analysis confirms 
that $k(L/B)$ exhibits two simultaneously present local 
maxima: a dominant peak near $L/B \approx 1.7$ (boundary 
interaction mode) and a secondary peak near $L/B \approx 
0.5$ (local Poisson mode), with the secondary reaching 
approximately $41$--$90\%$ of the dominant value as $\nu$ 
increases through this window. The grey shaded band in 
Fig.~\ref{fig:supp_mode_map_extended}(b) marks this 
two-mode coexistence window; the dashed grey line shows the 
secondary peak location within it. As $\nu$ increases 
through this window the local Poisson peak grows until the 
two peaks exchange dominance near $\nu \approx 0.28$, 
producing the discontinuous drop in $(L/B)^*$ visible in 
Fig.~\ref{fig:mode_map} of the main text.

Above the crossover ($\nu \gtrsim 0.28$), $(L/B)^*$ 
settles near $L/B \approx 0.45$. For $\nu \in [0.30, 
0.40]$, $k(L/B)$ is nearly insensitive to aspect ratio: 
finite-element results show that $k$ varies by less than 
5\% across $L/B \approx 0.45$--$0.90$ in this range, 
so $(L/B)^*$ represents an approximate centroid of a 
broad flat optimum rather than a sharp geometric optimum. 
Toward the incompressible limit ($\nu \to 0.45$), the 
peak sharpens and the 5\%-range narrows back toward 
$L/B \approx 0.30$--$0.50$, consistent with the growing 
dominance of the local Poisson mechanism as the bulk 
modulus diverges. In all cases above the crossover, the 
boundary interaction mode no longer appears as a distinct 
local maximum of $k(L/B)$.

The peak value of $k$ grows rapidly for $\nu \gtrsim 0.25$ 
(Fig.~\ref{fig:supp_mode_map_extended}a), reflecting the 
divergence of the plane-strain bulk modulus $E/(2(1-2\nu))$ 
as $\nu \to 0.5$. Numerical results are limited to 
$\nu \leq 0.45$ to avoid volumetric locking in the 
displacement-based finite element formulation.

\begin{figure}[htbp]
    \centering
    \includegraphics[width=\linewidth]{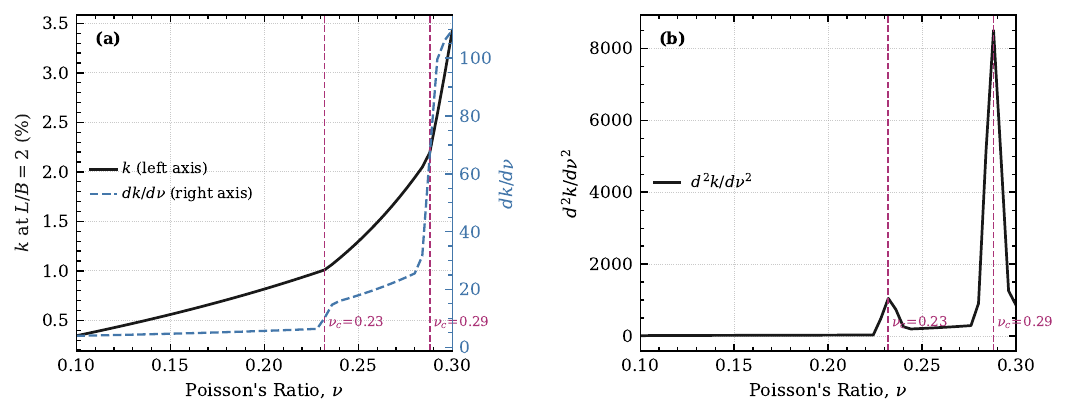}
    \caption{Rate of change of $k$ with $\nu$ at fixed 
    $L/B = 2$, from the same data as 
    Fig.~\ref{fig:bifurcation_analysis} (left panel).
    (\textit{a})~$k$ (left axis, solid) and $dk/d\nu$ 
    (right axis, dashed blue); purple dashed lines mark 
    kink locations $\nu_c$.
    (\textit{b})~$d^2k/d\nu^2$; peaks confirm the kinks 
    are genuine non-analyticities corresponding to the 
    bifurcation transitions in 
    Fig.~\ref{fig:bifurcation_analysis} (right panel).}
    \label{fig:supp_dk_dnu}
\end{figure}

\paragraph*{Rate of change of $k$ with $\nu$.}
Figure~\ref{fig:supp_dk_dnu} shows $dk/d\nu$ and 
$d^2k/d\nu^2$ at fixed $L/B = 2$, computed from the 
same finite-element data as Fig.~\ref{fig:bifurcation_analysis} 
(left panel) of the main text. The second derivative 
peaks at the kink locations $\nu_c$ identified in 
Fig.~\ref{fig:bifurcation_analysis}, confirming that 
these kinks are genuine non-analyticities in $k(\nu)$ 
rather than numerical noise. They correspond to the 
bifurcation points in the spatial stress topology 
(Fig.~\ref{fig:bifurcation_analysis}, right panel) 
and to the mode transition visible in 
Fig.~\ref{fig:mode_map}.

\begin{figure}[htbp]
    \centering
    \includegraphics[width=0.6\linewidth]{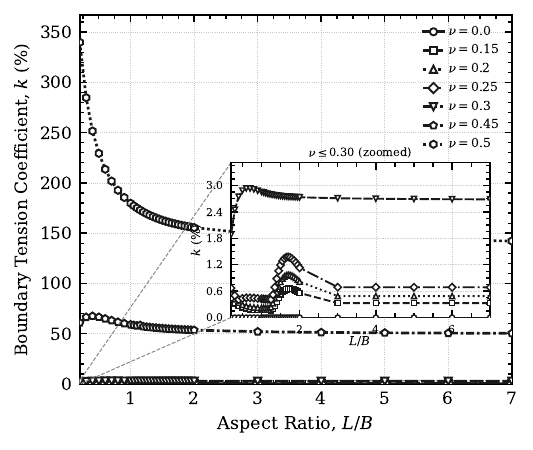}
    \caption{Boundary tension coefficient $k$ versus 
    $L/B$ extended to $\nu = 0.45$ and $\nu = 0.5$; 
    inset shows $\nu \leq 0.30$ at the scale of 
    Fig.~\ref{fig:k_vs_aspect_ratio}. The peak shifts 
    toward squat geometries and $k$ grows rapidly 
    as $\nu \to 0.5$. The response at $\nu = 0.5$ is computed via a mixed $P2/P1$ formulation to bypass volumetric locking.}
\end{figure}

\clearpage

\subsection*{Extension to Auxetic Materials ($\nu < 0$)}

To further validate that the central splitting failure mode is a specific
consequence of constrained lateral \textit{expansion}, we performed control
simulations on auxetic materials ($\nu < 0$). In this regime, global
compression induces lateral contraction. The clamped boundaries resist this
contraction, effectively ``pulling'' the specimen edges outward to maintain
the fixed width $B$.

\begin{figure}[htbp]
    \centering
    \includegraphics[width=0.3\linewidth]{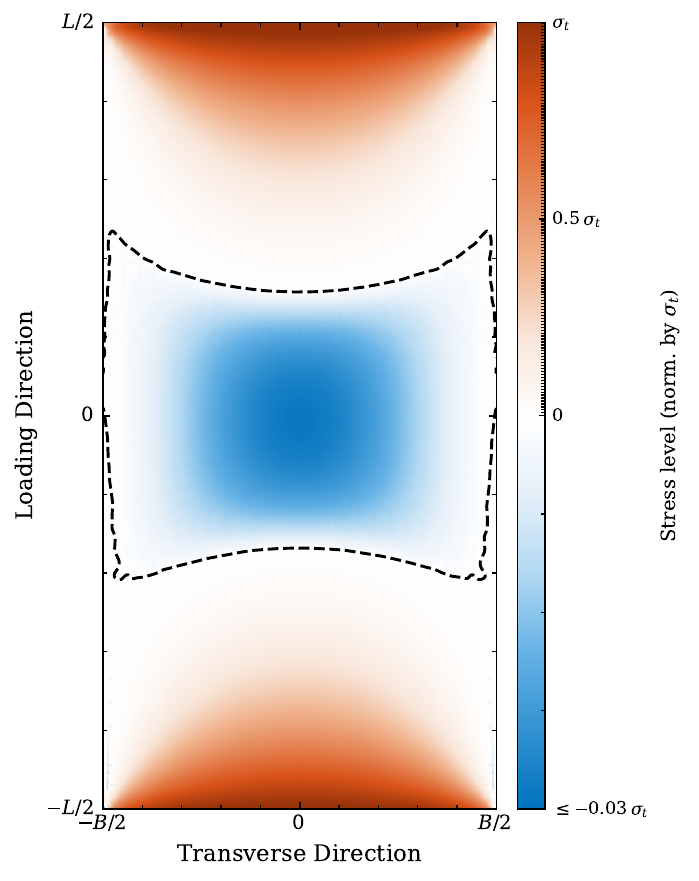}
    \caption{Transverse stress distribution ($\sigma_{xx}$) for an auxetic material ($\nu = -0.3$, $L/B=2.0$). Tension ($\sigma_{xx} > 0$, red) is generated at the clamped boundaries due to the restriction of lateral contraction, but decays rapidly towards the center.}
    \label{fig:supp_auxetic_stress}
\end{figure}

As illustrated in Fig.~\ref{fig:supp_auxetic_stress}, unlike the $\nu > 0$
case where boundary-induced tension localizes at the specimen center, the
transverse tension in auxetic materials is localized strictly at the clamped
boundaries. This manifests as a boundary stress concentration that decays
rapidly toward the center.

\begin{figure*}[htbp]
    \centering
    \begin{subfigure}[b]{0.45\textwidth}
        \centering
        \includegraphics[width=\textwidth]{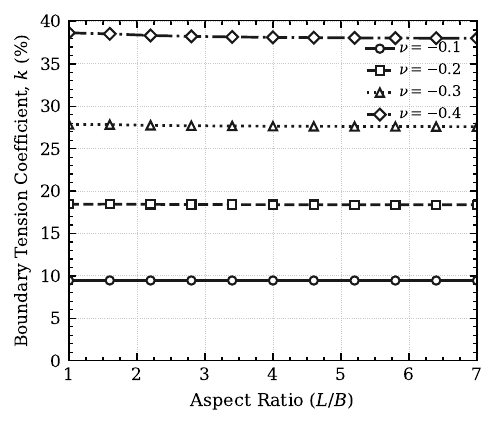}
        \caption{Monotonic decay with Aspect Ratio}
        \label{fig:auxetic_LB}
    \end{subfigure}
    \hfill
    \begin{subfigure}[b]{0.45\textwidth}
        \centering
        \includegraphics[width=\textwidth]{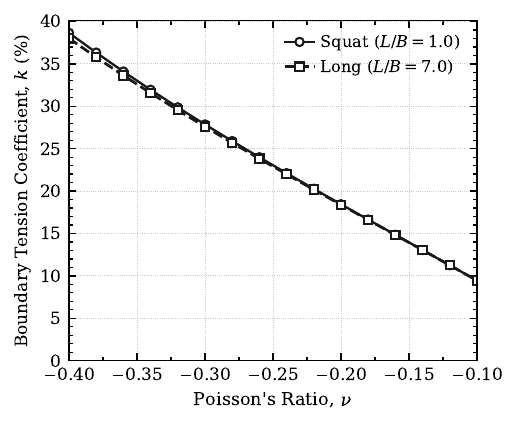}
        \caption{Linear scaling with Poisson's Ratio}
        \label{fig:auxetic_nu}
    \end{subfigure}
    \caption{Parametric analysis of the boundary tension coefficient $k$ for auxetic materials. (\textit{a})~Unlike the $\nu > 0$ case, $k$ is nearly independent of specimen aspect ratio across the full range $1 \leq L/B \leq 7$, confirming the absence of geometric mode transitions. (\textit{b})~$k$ scales linearly with $\nu$, and the squat ($L/B = 1.0$) and long ($L/B = 7.0$) specimens are nearly indistinguishable, consistent with the aspect-ratio independence shown in~(\textit{a}).}
    \label{fig:supp_auxetic_parametric}
\end{figure*}

Consequently, the boundary tension coefficient $k$ (here representing the
maximum tension at the clamped boundaries) exhibits no geometric mode
transitions and is essentially independent of specimen aspect ratio, as shown
in Fig.~\ref{fig:auxetic_LB}. This is in sharp contrast to the $\nu > 0$
case, where $k$ depends strongly and non-monotonically on $L/B$ through
competing boundary interaction and local Poisson modes. Instead, for auxetic
materials, $k$ is governed solely by the Poisson's ratio: as shown in
Fig.~\ref{fig:auxetic_nu}, $k$ scales linearly with $|\nu|$ and is
insensitive to geometry across the full range $1 \leq L/B \leq 7$. This
confirms that the symmetry-breaking pitchfork bifurcations in the spatial
topology of the boundary-induced tension field reported in the main text are
a unique feature of the kinematic constraint on lateral \textit{expansion},
absent when the constraint acts on lateral \textit{contraction}.

This stress distribution predicts a fundamental change in the failure
mechanism: tensile fracture in an auxetic material would initiate at the
clamped ends of the specimen rather than at the geometric center, because the
boundary-induced tensile stress concentration is localized there. However, a
competing failure mode must be noted. The global compressive load makes the
specimen susceptible to classical Euler buckling. For slender specimens, this
macroscopic structural instability would likely occur at a much lower
compressive load than that required to initiate fracture. Therefore, the
end-splitting failure mode predicted here would likely only be observable in
very squat auxetic specimens where Euler buckling is geometrically suppressed.

\clearpage
\subsection*{Experimental Data Parameters}

Table~\ref{tab:supp_materials} details the specific material properties, geometric aspect ratios, and literature sources used to generate the theoretical strength predictions and scale the experimental data presented in the main text (Fig.~6).

\begin{table*}[htbp]
  \centering
  \caption{Material parameters and sources for experimental data used for model validation.}
  \label{tab:supp_materials}
  \footnotesize 
  \setlength{\tabcolsep}{5pt} 

  \begin{tabular}{@{} ll c c c c p{0.35\textwidth} @{}} 
    \hline\hline 
    \textbf{Material Name} & \textbf{Citation} & \textbf{E (GPa)} & \textbf{$\nu$} & \textbf{$\sigma_t$ (MPa)} & \textbf{$L/B$} & \textbf{Description and Source Justification} \\
    \hline 

    Rock (Synthetic) & Tien et al., 2006~\cite{Tien2006} & 21.7  & 0.23  & 9.6   & 2.00 & Parameters correspond to  "Material A" (isotropic). $\sigma_t$ is Brazilian indirect tensile strength. Experimental data for $\alpha=0^\circ$ loading (axial splitting mode). \\
    \\[1ex] 

    Sintered AlN    & Chen \& Ravichandran, 1996~\cite{Chen1996} & 313.0 & 0.237 & 325.0 & 1.15 & Experimental data for quasi-static case. $\sigma_t$ is an assumed flexural strength. \\
    \\[1ex]

    Pyroceram 9606   & Lankford, 1995~\cite{Lankford1995}    & 119.3 & 0.245 & 138.0 & 2.00 & Elastic constants and $\sigma_t$ sourced from manufacturer datasheet~\cite{Corning1971}. Experimental data for quasi-static case. \\
    \\[1ex]

    SiC-N            & Lee et al., 2005~\cite{Lee2005}  & 465.0 & 0.16  & 580.0 & 2.00 & All model parameters are reported in the source text. \\
    \\[1ex]

    Westerly Granite & Brace et al., 1966~\cite{Brace1966} & 72.2  & 0.25  & 10.0  & 2.00 & Model parameters are assumed representative values for this material. \\
    \\[1ex]

    Rock Salt (NaCl) & Handin, 1953~\cite{Handin1953} & 36.9  & 0.253  & 1.5   & 2.00 & Model parameters are assumed representative values for this material. \\
    \hline\hline 
  \end{tabular}
\end{table*}

\subsection*{Scale separation and the ratio $d/B$}
Our framework treats the specimen as a homogeneous continuum, which requires the characteristic microstructural size $d$ (grain, crystallite, or constituent particle size) to be small compared to the length scale over which the boundary-induced stress field varies. Because our failure criterion is evaluated at a smooth interior maximum of the pre-fracture stress field, rather than at a crack tip, the relevant comparison length is not a crack-tip process zone but the scale of the tensile hotspot, which spans a finite fraction of the specimen width $B$ (Fig.~\ref{fig:fem_results}). 
The operative condition is therefore $d/B \lesssim 10^{-1}$, so that the specimen contains enough microstructural units for its response to be statistically
representative---the standard requirement for representing a heterogeneous solid as a homogeneous continuum~\cite{OstojaStarzewski2006PEM, OstojaStarzewski2007IJMCE}.
Table~\ref{tab:supp_scale} lists $d$ and $d/B$ for the materials of Fig.~\ref{fig:model_vs_experiment_scaled_regression} for which the microstructural scale is available in the source literature; the ratios range from $\sim\!10^{-4}$ for the fine-grained ceramics to $\sim\!10^{-2}$ for Westerly granite, in all cases at least an order of magnitude below this bound. For Westerly granite the specimen dimensions are not reported in the original source~\cite{Brace1966}, so its ratio is reliable only to order of magnitude. For rock salt (Handin 1953~\cite{Handin1953}) the specimen geometry and grain size are not reported in the accessible sources and are omitted here. 
Where $d/B$ is not negligible, the microstructure and the boundary-induced stress field are expected to interact, and capturing that interplay is a natural extension of the present continuum treatment.

\begin{table*}[htbp]
  \centering
  \caption{Characteristic microstructural length scale $d$ and the
  scale-separation ratio $d/B$ for the materials of Fig.~\ref{fig:model_vs_experiment_scaled_regression} for which the microstructural scale is available in the literature. $d$ is the mean grain, crystallite, or constituent particle size unless noted; $B$ is the specimen width (diameter, except where a gauge-section width applies).}
  \label{tab:supp_scale}
  \footnotesize
  \setlength{\tabcolsep}{5pt}

  \begin{tabular}{@{} ll c c c p{0.30\textwidth} @{}}
    \hline\hline
    \textbf{Material Name} & \textbf{Citation} & \textbf{$B$ (mm)}
    & \textbf{$d$} & \textbf{$d/B$} & \textbf{Basis for $d$} \\
    \hline

    Rock (Synthetic) & Tien et al., 2006~\cite{Tien2006} & 50.0
    & $\sim10^{-3}$~mm & $\sim3\times10^{-4}$
    & Constituent size not reported in the source. Material~A is a cement-based simulated rock; taking the constituent particle scale as $O(10$--$100~\mu$m) gives $d/B \sim 10^{-4}$--$10^{-3}$.  \\
    \\[1ex]

    Sintered AlN & Chen \& Ravichandran, 1996~\cite{Chen1996} & 4.76
    & $2$--$10~\mu$m & $\sim\!10^{-3}$
    & Grain size not reported. Microcracks along grain boundaries span
    $2$--$10~\mu$m~\cite{Chen1995Thesis},
    bounding the grain scale to this order; grains are comparably sized
    or finer. \\
    \\[1ex]

    Pyroceram 9606 & Lankford, 1995~\cite{Lankford1995} & 6.25
    & $0.5$--$2.0~\mu$m & $\sim\!3\times10^{-4}$
    & Grain size $0.5$--$2.0~\mu$m reported directly; $d/B$ evaluated at the coarse end ($2.0~\mu$m).\\
    \\[1ex]

    SiC-N & Lee et al., 2005~\cite{Lee2005} & 12.7 & $4~\mu$m ($1$--$8~\mu$m) & $\sim3\times10^{-4}$ & Mean grain size reported as $4~\mu$m (range $1$--$8~\mu$m). \\
    \\[1ex]

    Westerly Granite & Brace et al., 1966~\cite{Brace1966} &
    -- & $0.75$~mm & $\sim\!10^{-2}$ &
    Grain size is the representative value for Westerly granite (Tullis \& Yund 1977~\cite{TullisYund1977}). Specimen dimensions not reported in Brace 1966~\cite{Brace1966}; a representative gauge diameter gives $d/B$ of order $10^{-2}$. Brace used dogbone specimens
to suppress end-effect vertical splitting, so the plotted strengths
are faulting strengths and this row validates the strength--confinement
relation rather than the splitting morphology.\\
    \hline\hline
  \end{tabular}
\end{table*}

\subsection*{Comparison with Energetic Models}

\begin{figure}[ht!]
    \centering
    \includegraphics[width=0.48\columnwidth]{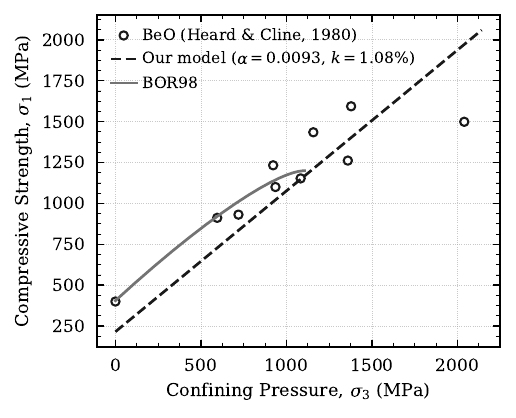}
    \caption{
    Comparison of the predicted compressive strength for beryllium oxide (BeO). 
    Solid line: energetic bifurcation model (BOR98)~\cite{Bhattacharya1998}.
    Dashed line: present boundary-driven model.}
    \label{fig:BeO_comparison_BOR98}
\end{figure}

Figure~\ref{fig:BeO_comparison_BOR98} compares our boundary-driven model with the energetic bifurcation model (BOR98)~\cite{Bhattacharya1998}. The experimental data for beryllium oxide (BeO) are taken from the compression tests of Heard and Cline (1980)~\cite{Heard1980}. Both formulations predict a broadly linear increase in compressive strength with confinement; however, they are founded on fundamentally different physical principles and domains of validity.

The BOR98 model is explicitly \textit{energetic} and \textit{mode-specific}. It treats the onset of axial splitting as a bifurcation between an intact and a notionally split configuration, determined by an energy balance between the bulk strain-energy release and the cost of creating new fracture surfaces. The stability condition depends on the parameter $\gamma/d$ (surface-energy density over a representative grain size) and on the contrast between the effective elastic moduli of the intact and notionally split phases ($E_u,\nu_u$ versus $E_s,\nu_s$), the latter often approximated using the Reuss bound (see Table~\ref{tab:supp_beo_params}). 

\begin{table}[h]
  \centering
  \caption{Material parameters used for the BeO comparison (Fig.~\ref{fig:BeO_comparison_BOR98}).}
  \label{tab:supp_beo_params}
  \footnotesize 
  \begin{tabular}{@{} l c c c c @{}} 
    \hline\hline
    \textbf{Parameter/Phase} & \textbf{E (GPa)} & \textbf{$\nu$} & \textbf{$\sigma_t$ (MPa)} & \textbf{$\boldsymbol{\gamma/d}$ (GJ/m$^3$)} \\
    \hline
    \textbf{The current model} & 394.0 & 0.2370 & 250.0 & N/A \\
    \hline
    \textbf{BOR98 Intact Phase ($E_u, \nu_u$)} & 394.0 & 0.2370 & N/A & \multirow{2}{*}{$2.652 \times 10^{-7}$} \\
    \textbf{BOR98 Split Phase ($E_s, \nu_s$)} & 393.0 & 0.2376 & N/A & \\
    \hline\hline
  \end{tabular}
\end{table}

The parameters in Table~\ref{tab:supp_beo_params} differ slightly from those listed in BOR98 Table~2: the values published in the original paper do not reproduce the authors' own Fig.~2, suggesting a transcription inconsistency in that table. The parameters used here are those that correctly reproduce the BOR98 theoretical curve shown in Fig.~\ref{fig:BeO_comparison_BOR98}.

As confinement increases, the elastic energy available for release decreases, and above a critical pressure the energetic criterion for splitting can no longer be met. BOR98 explicitly notes that beyond this point “the observed failure changes from axial splitting to crushing and shear-dominated compaction,” and their theoretical curve therefore \textit{terminates} at that confinement limit. 

Both frameworks require a single fitted parameter: the present model fits $\alpha$ to the experimental strength data, while BOR98 
fits $\gamma/d$ (surface-energy density per representative grain size). However, BOR98 additionally requires prescribing the elastic 
moduli of both the intact phase ($E_u$, $\nu_u$, estimated as the Voigt--Reuss average of single-crystal constants) and the notionally split phase ($E_s$, $\nu_s$, taken as the Reuss lower bound). 
The present model does not require these parameters for the split phase; rather, the strength--confinement relation emerges directly from the superposition of the boundary-induced transverse tension and the applied confinement, parameterized solely by macroscopic continuum properties ($E$, $\nu$, $\sigma_t$) and specimen geometry ($L/B$). 
This formulation predicts a continuous, linear strengthening trend without invoking separate regimes or energetic bifurcations, and remains valid as confinement increases and fracture morphology transitions from splitting to crushing.

Qualitatively, both models reproduce the experimental trends for high-strength ceramics such as AlN and Al$_2$O$_3$. However, for materials like limestone and, most notably, BeO, the distinction becomes clear. The BOR98 prediction ceases near $\sigma_3 \!\approx\! 1$ GPa, whereas the experimental data of Heard and Cline~\cite{Heard1980} continue to show strengthening well beyond this range. Our scaled model accurately follows these data into the high-confinement regime, capturing the apparent ``crushing'' behavior without additional parameters or a regime switch.

\begin{figure*}[htbp]
    \centering
    \begin{subfigure}[b]{0.48\textwidth}
        \centering
        \includegraphics[width=\textwidth]{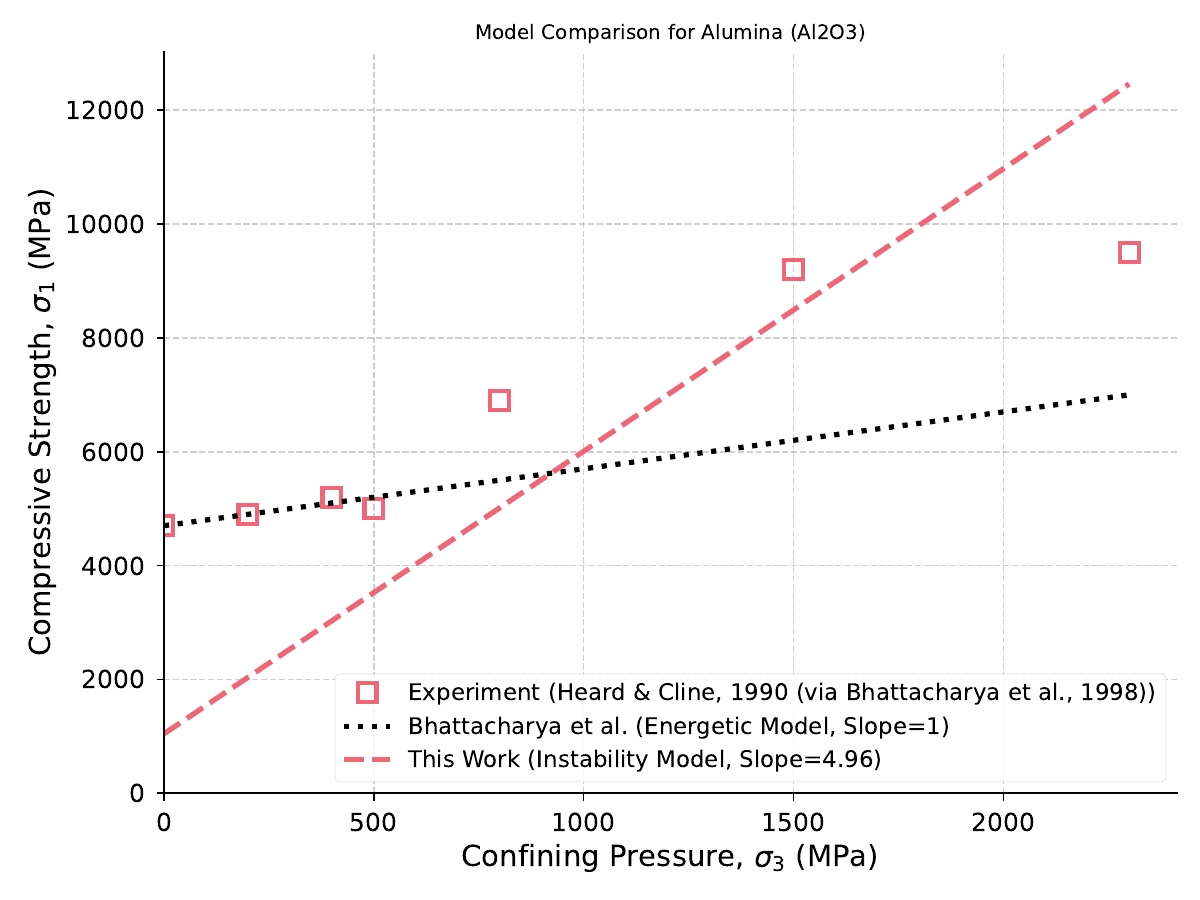} 
        \caption{Alumina (Al$_2$O$_3$)}
        \label{fig:supp_al2o3}
    \end{subfigure}
    \begin{subfigure}[b]{0.48\textwidth}
        \centering
        \includegraphics[width=\textwidth]{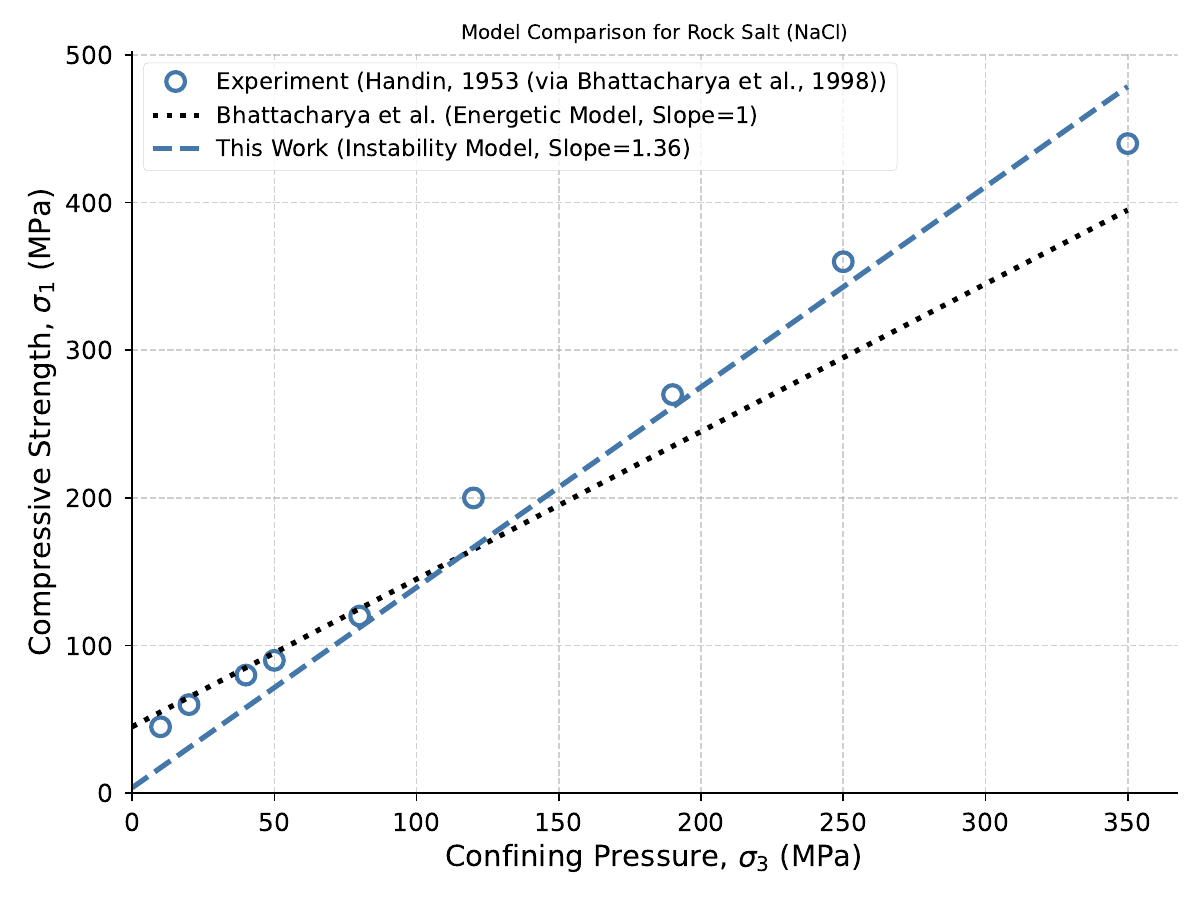} 
        \caption{Rock Salt (NaCl)}
        \label{fig:supp_nacl}
    \end{subfigure}
    \caption{Comparison of the predicted compressive strength for (a) Alumina, and (b) Rock Salt. Dashed lines represent our scaled model; solid lines represent the energetic bifurcation model (BOR98)~\cite{Bhattacharya1998}. Experimental data are taken from~\cite{Heard1980} (Al$_2$O$_3$) and ~\cite{Handin1953} (NaCl).}
    \label{fig:supp_bor98_comparisons}
\end{figure*}

To further demonstrate the robustness of this framework, Figure~\ref{fig:supp_bor98_comparisons} presents comparisons between our boundary-induced model and the energetic bifurcation model (BOR98)~\cite{Bhattacharya1998} for two additional materials: Alumina (Al$_2$O$_3$) and Rock Salt (NaCl). In both cases, our model (dashed lines) successfully captures the strengthening trend across the full range of confinement, consistently outperforming the energetic model's termination limits.

\end{document}